\documentstyle [12pt]{article}

\begin{document}

\begin{titlepage}

\title{Quantum mechanics and elements of reality 
inferred from joint measurements\footnote{J. Phys. A: 
Math. Gen. {\bf 30}, 725-732 (1997).}}

\author{Ad\'{a}n Cabello\thanks{Present 
address: {\em Departamento de F\'{\i}sica Aplicada, Universidad de Sevilla,
41012 Sevilla, Spain.} 
Electronic address: fite1z1@sis.ucm.es} 
\and
Guillermo Garc\'{\i}a-Alcaine\thanks{Electronic address: 
fite114@sis.ucm.es}\\
{\em Departamento de F\'{\i}sica Te\'{o}rica,}\\
{\em Universidad Complutense, 28040 Madrid, Spain.}}

\date{\today}

\maketitle


\begin{abstract}
The Einstein-Podolsky-Rosen argument on quantum mechanics 
incompleteness is formulated in terms of elements of reality 
inferred from {\em joint} (as opposed to {\em alternative}) measurements, 
in two examples involving entangled states of 
three spin-$\frac{1}{2}$ particles. 
The same states allow us to obtain proofs of the incompatibility 
between quantum mechanics and elements of reality.\\
\\
PACS numbers: 03.65.Bz

\end{abstract}

\end{titlepage}

\section{Introduction}
The layout of the paper is as follows. In section 2 we briefly review 
some opinions on a controversial point in Einstein-Podolsky-Rosen's 
(EPR) \cite{EPR35} argument. In section 3 we introduce some notations 
and a more detailed form of the sufficient condition for joint inference 
of several elements of reality (ERs) in the same individual system. 
In section 4 we formulate EPR's incompleteness argument in terms of ERs 
inferred from joint measurements (instead of alternative incompatible 
measurements as in the original paper and all its sequels), in two entangled 
states of three spin-$\frac{1}{2}$ particles: an extension to three particles 
of Hardy's two-particle state \cite{Hardy93}, and GHZ-Mermin state 
\cite{GHZ89,Mermin90a}. In section 5 we show the incompatibility between 
quantum mechanics (QM) and ERs (Bell's theorem), in the same entangled 
states used in section 4. Finally we summarize our conclusions 
in section 6.

\section{Elements of reality and EPR's incompleteness argument}
	In 1935, EPR presented an argument to prove that QM only provides an 
``incomplete'' description of physical reality \cite{EPR35}. 
Their conclusions were striking because of ``the very mild character 
of the sufficient condition for the reality of a physical quantity 
on which their argument hinged'' \cite{Mermin90b}:
``If, without in any way disturbing a system, we can predict with certainty 
(i.e., with probability equal to unity) the value of a physical quantity, 
then there exists an element of physical reality corresponding to this 
physical quantity.''

There is a problem, however, in EPR's argument: ``... one cannot make both 
measurements [position or momentum of one of the particles in the original 
EPR example], and hence both predictions [position or momentum of the other 
particle], simultaneously'' \cite{HB92}. The joint (``simultaneous'') 
existence in the same individual system of ERs corresponding to two 
incompatible observables is inferred from the {\em possibility} of measuring 
either of two also mutually incompatible observables\footnote{
In \cite{Redhead87}, Redhead discusses how, if the {\em minimal 
instrumentalistic interpretation} of the QM formalism is complemented by any 
of several possible {\em views} (p 45), then ``the argument for incompleteness 
of the QM goes through without any consideration of the alternative 
possibilities of measuring [several incompatible observables]'' (p 78). 
The locality principle and the inference of only one element of reality 
(in the singlet state, in Redhead's example) will be enough to prove QM 
incompleteness (p 77). In this paper we will prove how EPR's argument on 
QM incompleteness can be reformulated restricting ourselves to a minimal 
set of assumptions, both on QM interpretation ({\em minimal instrumentalistic 
interpretation}, in Redhead's terminology), and on the condition for 
existence of ERs.}; this fact has been pointed out as ``the most significant 
lacuna'' in EPR's reasoning \cite{Howard85}. For EPR's incompleteness 
argument to be valid \cite{HB92,Howard85}, {\em counterfactual definiteness} 
(CFD) \cite{HB92,Ballentine90} is required, either as an additional 
assumption \cite{Howard85}, or deduced from locality \cite{Ballentine90}. 
Independently of how natural it may appear, in our opinion this spoils EPR's 
own dictum: ``The elements of the physical reality cannot be determined by 
{\em a priori} philosophical considerations, but must be found by an appeal to 
results of experiments and measurements'' \cite{EPR35}.

In their argument, EPR use the ambiguous expression ``can predict'' in the 
condition for existence of ERs in a broad or ``weak'' sense 
\cite{HB92,CS78,GHSZ90,Stapp91,d'Espagnat93,Shimony93}, meaning that it is 
{\em possible} to make any of the {\em alternative} measurements that 
would provide the 
data for either inference. But, as it has been pointed out 
\cite{HB92,CS78,GHSZ90,Stapp91,d'Espagnat93,Shimony93}, ``can predict'' can 
also be interpreted in a narrow or ``strong'' sense, meaning that we 
actually {\em do have} sufficient data to predict with certainty 
the {\em concrete} 
result to which the ER is ascribed. It is a common belief that EPR's 
argument does not work if this strong sense is required: 
``The EPR argument goes through only if ``can predict'' is understood in 
the weak sense'' \cite{Shimony93} (see also p 1885 of \cite{CS78}, 
or p 142 of \cite{d'Espagnat93}); EPR themselves said: ``Indeed, one would 
not arrive at our conclusion if one insisted that two or more physical 
quantities can be regarded as simultaneous elements of reality {\em only 
when they can be simultaneously measured or predicted}. On this point 
of view, since either one or the other, but not both simultaneously, 
of the quantities {\em P} and {\em Q} can be predicted, they are 
not simultaneously 
real. This makes the reality of {\em P} and {\em Q} depend upon 
the process of measurement carried out on the first system, which 
does not disturb the second system in any way. No reasonable 
definition of reality could be expected to permit this'' \cite{EPR35}. 
The main aim of this paper is to show how we can formulate an EPR argument 
using ERs predicted in the strong sense\footnote{Such form of the argument 
will not work for systems with only two spacelike separated parts: in 
these systems, any couple of joint ERs predicted in the strong sense will 
correspond to one observable on each of the two spacelike separated parts, 
that are always compatible; in order to have ERs in the strong sense for 
incompatible observables we need three or more spacelike separated parts.}.

\section{Elements of reality inferred from joint measurements}
We will call {\em strong elements of reality} (SERs) those obtained from 
EPR's criterion, taking ``can predict'' in the strong sense. 
The name also reflects their {\em actuality}, in contradistinction with the 
ones used in EPR's paper and in most of the literature: SERs have definite 
predicted values, instead of an abstract existence without concrete 
values as the ERs inferred using the weak sense of ``can predict''. 
Note, however, that EPR's criterion is a {\em sufficient} condition, and 
that the set of assumptions used to infer SERs is in fact {\em weaker}: 
the CFD implicitly used in the original EPR argument is not 
required\footnote{The word {\em counterfactual} is used in several 
contexts: quoting Ballentine \cite{Ballentine90}, ``{\em Counterfactual 
definiteness} (abbreviated CFD) ...occurs in the EPR argument when they 
assert that if we had measured the position of particle 1 we could have 
learned the position $x_2$ of particle 2, and if we had measured the 
momentum of particle 1 we could have learned the momentum $p_2$ of 
particle 2. Although only one of these measurements can actually be 
carried out in a single case, the conclusion that both values $x_2$ and 
$p_2$ are well defined in nature is an instance of CFD.'' 
{\em Counterfactual} is used also to qualify those predictions that are 
not actually tested (or even that {\em cannot} be tested); in this sense, 
we subscribe Pitowsky's opinion \cite{Pitowsky91}: ``This [the fact that 
the experiment to test a prediction cannot be conducted] does not render 
the prediction invalid. It simply makes it untestable.'' We think that 
counterfactuality in this second sense is less worrying that the one 
involved in the first, where the {\em joint} inference of several ERs 
is based on {\em alternative} incompatible measurements.}.

We will guarantee the clause ``without in any way disturbing a system'' in 
EPR's sufficient condition by requiring that the system to which the ER is 
assigned be outside the future light cones of the points at which we perform 
any observation needed to make the prediction. This is how the ERs were 
inferred in EPR's original paper: ``since at the time of measurement the two 
systems no longer interact, no real change can take place in the second 
system in consequence of anything that may be done to the first system'' 
\cite{EPR35}. More explicitly, we formulate the sufficient condition for 
existence of a SER in the following way.

Let us consider a physical system with two parts $S_1$, $S_2$, and two 
spacelike separated regions $R_1$, $R_2$  of the respective world tubes. 
If we can predict with certainty the concrete value of a physical quantity 
in $R_2$ from the result of a measurement performed in $R_1$, then there is 
a {\em strong element of reality} corresponding to that physical quantity, 
{\em at least} in the part of the world tube of $S_2$ outside the future 
light cones with vertices in $R_1$ and prior to any external perturbation 
of $S_2$.

The SER need not be the value of the physical quantity itself, as long as 
it determines univocally such value. For instance, in Bohm's theory the 
position of a particle in a Stern-Gerlach apparatus determines the value of 
the corresponding spin component \cite{BH93}. Nevertheless, to simplify 
the discussion we will identify the SER with the value of the physical 
quantity that it predicts.

Our next step is to illustrate when several SERs can be jointly assigned 
to the same individual system. Suppose that we make a measurement in a 
region $R_1$ of the world tube of $S_1$, and that the result obtained allows 
us to infer a SER $r^{(2)}$ for $S_2$ following our previous criterion. 
On the same individual system, suppose that a second observer makes a 
measurement in a region $R_2$ of the world tube of $S_2$, spacelike 
separated from $R_1$, and that his result enables him to infer a SER 
$r^{(1)}$ for $S_1$ in its originally prepared state (prior to any 
disturbance). The persistence of the SER $r^{(2)}$ after the measurement 
at $R_2$ is not guaranteed (in general the measurement disturbs the 
subsystem $S_2$; it could even destroy it!); the same can be said about 
the SER $r^{(1)}$ after the measurement at $R_1$. According to the 
sufficient condition for existence of SERs, we can assign to that individual 
system two {\em joint} SERs, $r^{(1)}$ and $r^{(2)}$, at least in the 
parts of the world tubes outside the future light cones with vertices 
in $R_1$ and $R_2$ (loosely speaking, in the parts of the world tubes 
``previous'' to the measurements).

For any pair of (point-like) events, one in each region, there is always an 
inertial reference frame in which both events are simultaneous. 
Therefore, it is tempting to adopt the usual terminology and denote the 
joint SERs as ``simultaneous'' SERs, but we have avoided this in order to 
emphasize that the combined existence of $r^{(1)}$ and $r^{(2)}$ in the same 
individual physical system does not depend on any external choice of a 
reference frame. We need not worry about instantaneity of measurements 
or simultaneity of three-dimensional sections of the four-dimensional 
world tubes of $S_1$ and $S_2$, and the generalization to systems with 
three or more extended spacelike separated parts is immediate: the only 
requisite on the compound system is that we can make predictions with 
certainty about properties (possibly non-local) of the individual system in 
some spacetime regions (prior to any external disturbance), based on 
measurements performed in other spacelike separated regions.

\section{EPR's argument in terms of joint SER}
We are going to follow EPR's previously quoted dictum on the necessity of 
finding ERs ``by an appeal to results of experiments'', and reformulate 
their argument using only ERs jointly inferred from actual measurements 
(our previously defined joint SERs), without {\em a priori} use of 
alternative inferences. Joint SERs will be obtained for observables that have 
{\em no} common eigenstate; this is a stronger condition than the simple 
incompatibility; even if they do not have a basis of common eigenstates, two 
incompatible observables can share some eigenstates; the existence of joint 
SERs in one of these states will not prove QM incompleteness (of course,
$\hat P$, $\hat Q$ have no common eigenstate, but other incompatible 
observables without this property have been used in the literature).

Consider a system of three spin-$\frac{1}{2}$ particles in spacelike 
separated regions, prepared in the entangled spin state (in the basis of 
eigenstates of the operators $\hat \sigma _z^{(j)}$, $j=1, 2, 3$),
\begin{equation}
\left| \psi  \right\rangle = a \left( {\left| {+++} \right\rangle -
\left| {+-+} \right\rangle -\left| {-++} \right\rangle } \right) + 
b \left| {---} \right\rangle\,,
\label{uno}
\end{equation}
with $3\left| a \right|^2  +\left| b \right|^2 = 1$, and $ab\ne 0$.

In this state,
\begin{equation}
P_\psi \left( {\sigma _x^{(2)}=-1 \left|\,{\sigma _z^{(1)}=+1} \right.} 
\right) = 1\,,
\label{dos}
\end{equation}
\begin{equation}
P_\psi \left( {\sigma _x^{(1)}=-1 \left|\,{\sigma _z^{(2)}=+1} \right.} 
\right) = 1\,,
\label{tres}
\end{equation}
where $P_\psi \left( {\sigma _x^{(2)}=-1 \left|\,{\sigma _z^{(1)}=+1} \right.} 
\right)$ denotes the conditional probability of obtaining the value 
$\sigma _x^{(2)}=-1$ if $\sigma _z^{(1)}=+1$. If we measure 
$\hat \sigma _z^{(1)}$ on the first particle and obtain the result $+1$, 
property (\ref{dos}) allows us to assign a SER $\sigma _x^{(2)}=-1$ to the 
second particle, at least as long as it is not disturbed. Analogously, 
if a second observer measures $\hat \sigma _z^{(2)}$ on the second 
particle and 
obtains the result $+1$, property (\ref{tres}) allows him to assign a SER 
$\sigma _x^{(1)}=-1$ to the first unperturbed particle\footnote{Of course, 
after measuring  $\sigma _z^{(1)}$, $\sigma _z^{(2)}$, the values predicted 
for $\sigma _x^{(1)}$, $\sigma _x^{(2)}$ cannot be verified, and in this 
sense these joint SERs are also counterfactual, but at least the 
measurements needed to infer both values can be made in the same individual 
system, and joint counterfactual (alternative) inferences are not involved; 
see footnote 3.}.

Consider now the non-local (and non-factorizable) observable defined by the 
projector
\begin{equation}
\hat \pi ^{(1+2)} = 1 -\left| {--} \right\rangle \left\langle {--} \right|\,,
\label{cuatro}
\end{equation}
in the system formed by the first and second particles. This observable 
$\hat \pi ^{(1+2)}$ is compatible with $\hat \sigma _z^{(1)}$ and 
$\hat \sigma _z^{(2)}$ but not with $\hat \sigma _x^{(1)}$ and 
$\hat \sigma _x^{(2)}$; in fact, it is easy to check that there is no common 
eigenstate to the three operators $\hat \sigma _x^{(1)}\otimes 1^{(2)}$, 
$1^{(1)}\otimes \hat \sigma _x^{(2)}$ and $\hat \pi ^{(1+2)}$.

In state (\ref{uno}),
\begin{equation}
P_\psi \left( {\pi ^{(1+2)}=1 \left|\,{\sigma _z^{(3)}=+1} \right.} 
\right) = 1\,.
\label{cinco}
\end{equation}
Consequently, if a third observer measures $\hat \sigma _z^{(3)}$ 
and obtains the 
result $+1$, he can assign a SER $\pi ^{(1+2)}=1$ to the system formed by 
the first and second particles.

The probability of obtaining the results $\sigma _z^{(1)}=+1$, 
$\sigma _z^{(2)}=+1$, $\sigma _z^{(3)}=+1$ in a joint measurement of the 
three observables $\hat \sigma _z^{(j)}$ in spacelike separated regions of 
the same individual physical system is not zero:
\begin{equation}
P_\psi \left( {\sigma _z^{(1)}=+1,\sigma _z^{(2)}=+1,\sigma _z^{(3)}=+1} 
\right)=\left| a \right|^2\,.
\label{seis}
\end{equation}
In an individual physical system in which these three results are obtained, 
following (\ref{dos}), (\ref{tres}) and (\ref{cinco}) we can infer three 
joint SERs: $\sigma _x^{(1)}=-1$, $\sigma _x^{(2)}=-1$ and $\pi ^{(1+2)}=1$. 
Because there is {\em no} common eigenstate of the corresponding observables, 
according to EPR we would conclude that the quantum state ``does not provide 
a complete description of the physical reality'' \cite{EPR35} of this 
individual system.

We could attempt a similar incompleteness argument in the state of the first 
two particles given by
\begin{equation}
\left| \eta  \right\rangle ={1 \over {\sqrt 3}}\left( {\left| {++} 
\right\rangle -\left| {+-} \right\rangle -\left| {-+} 
\right\rangle } \right)\,,
\label{siete}
\end{equation}
which is an example of a Hardy state \cite{Hardy93} that verifies properties 
(\ref{dos}) and (\ref{tres}) and is an eigenstate of $\hat \pi ^{(1+2)}$ 
with eigenvalue 1. But then, the values $\sigma _x^{(1)}=-1$, 
$\sigma _x^{(2)}=-1$ and $\pi ^{(1+2)}=1$ would not satisfy our condition 
to be considered joint SERs, because the preparation of the state 
(\ref{siete}), in which the value $\pi ^{(1+2)}=1$ rest, is not spacelike 
separated from the measurements used to infer the two other values; the 
measurements of $\hat \sigma _z^{(1)}$, $\hat \sigma _z^{(2)}$ are in the 
future of the preparation of the state (\ref{siete}), and therefore their 
precise results in an individual system could be influenced by the 
preparation.

As we said in footnote 2, three or more spacelike separated parts are needed 
in order to meet the condition for existence of joint SERs for incompatible 
observables (of course, ERs might exist without this {\em sufficient} 
condition being fulfilled, but following EPR, we should not concern 
ourselves with more precise definitions of the ERs). The third particle in 
our example (\ref{uno}) is a device to allow us the use of the sufficient 
condition for existence of joint SERs, although the SERs inferred involve 
only the first two particles; in the next example the three particles play 
a more symmetrical role.

A stronger incompleteness proof can be worked out in the GHZ-Mermin 
\cite{GHZ89,Mermin90a} state of three spin-$\frac{1}{2}$ particles:
\begin{equation}
\left| \mu  \right\rangle ={1 \over {\sqrt 2}}\left( {\left| {+++} 
\right\rangle -\left| {---} \right\rangle } \right)\,.
\label{ocho}
\end{equation}
Let us denote 
$\hat A_1=\hat \sigma _x^{(2)}\otimes \hat \sigma _y^{(3)}$, 
$\hat A_2=\hat \sigma _y^{(1)}\otimes \hat \sigma _x^{(3)}$, 
$\hat A_3=\hat \sigma _x^{(1)}\otimes \hat \sigma _y^{(2)}$. 
In this state we have,
\begin{equation}
P_\mu \left( {A_j=\varepsilon _j\left|\,{\sigma _y^{(j)}=\varepsilon _j} 
\right.} \right)=1\,,\;\;\;\;\varepsilon _j=\pm 1\,,\;\;j=1,2,3\,.
\label{nueve}
\end{equation}

There is no common eigenstate to any two of the observables $\hat A_j$, 
and nevertheless we can infer joint SER for the three of them, by measuring 
$\hat \sigma _y^{(j)}$ in spacelike separated regions, whatever the results 
obtained in these measurements (in our previous example we have to restrict 
ourselves to those individual physical systems in which $\sigma _y^{(j)}=+1$,
$j=1,2,3$). The price to pay for this extension of the argument to all 
individual systems in state (\ref{ocho}) is that the three incompatible 
observables to which the joint SER are assigned are now non-local (although 
factorizable), instead of one non-local and two local observables in 
the previous ``probabilistic'' example.

In both examples (\ref{uno}), (\ref{ocho}) the incompleteness argument has 
been formulated in terms of joint SERs, inferred from joint 
(``simultaneous'', in the usual terminology) local measurements in causally 
non-connected regions of the same individual system, which was considered 
impossible until now \cite{CS78,d'Espagnat93,Shimony93}.

\section{Bell-EPR theorem in terms of joint SER}
In recent years there have been many proofs of Bell-EPR theorem \cite{Bell64} 
``without inequalities'' \cite{Hardy93,GHZ89,Mermin90a,GHSZ90}. 
Arguments similar to those in the last section, in the same entangled states, 
lead us to proofs of these kind that are simple variants, using joint SERs, 
of Hardy's \cite{Hardy93} and Mermin's \cite{Mermin90a} results.

State (\ref{uno}) has the following additional properties
\begin{equation}
P_\psi \left( {\sigma _z^{(2)}=-1 \left|\,{\sigma _x^{(1)}=+1} \right.} 
\right) = 1\,,
\label{diez}
\end{equation}
\begin{equation}
P_\psi \left( {\sigma _z^{(1)}=-1 \left|\,{\sigma _x^{(2)}=+1} \right.} 
\right) = 1\,,
\label{once}
\end{equation}
\begin{equation}
P_\psi \left( {\sigma _x^{(1)}=+1,\sigma _x^{(2)}=+1,\sigma _z^{(3)}=+1} 
\right)=\frac{1}{4} \left| a \right|^2\,.
\label{doce}
\end{equation}
If we measure the observables $\hat \sigma _x^{(1)}$, $\hat \sigma _x^{(2)}$ 
and $\hat \sigma _z^{(3)}$, and obtain $+1$ in the three cases (the 
probability for this is not zero, according to (\ref{doce})), properties 
(\ref{diez}) and (\ref{once}), together with property (\ref{cinco}), allow 
us to infer respectively three joint SERs associated with the {\em compatible} 
observables $\hat \sigma _z^{(2)}$, $\hat \sigma _z^{(1)}$ and 
$\hat \pi^ {(1+2)}$, with predicted values $-1$, $-1$ and 1, respectively. 
But according to QM, in a joint measurement (feasible in principle) of 
these observables in any state, if the first two results are 
$\sigma _z^{(1)} = \sigma _z^{(2)} = -1$, the third one will necessarily 
be $\pi^ {(1+2)} = 0$: QM is not compatible with EPR's elements of reality, 
even in their less controversial form (joint SERs).

State (\ref{uno}) is the tensor product of the Hardy state (\ref{siete}) 
by state 
$\left| + \right\rangle $ of the third particle, entangled with the product 
of state $\left| {--} \right\rangle $ of the first two particles by state 
$\left| - \right\rangle $ of the third one. The presence of the third 
particle allows us to prove the incompatibility with QM (Bell's theorem) 
using three joint SERs that can never be found as results of the 
corresponding measurements, not only on state (\ref{uno}), but on {\em any} 
quantum state; by contrast, Hardy's two ERs, $\sigma _z^{(1)} = -1$, 
$\sigma _z^{(2)} = -1$, can never be obtained in a joint measurement in 
state (\ref{siete}), but states in which these results can be obtained do 
trivially exist.

The contradiction between ERs for compatible observables and QM found by 
Mermin \cite{Mermin90a} in the GHZ-Mermin state (\ref{ocho}) can also be 
formulated in terms of joint SERs. Let us denote 
$\hat B_1=\hat \sigma _y^{(2)}\otimes \hat \sigma _y^{(3)}$, 
$\hat B_2=\hat \sigma _y^{(1)}\otimes \hat \sigma _y^{(3)}$, 
$\hat B_3=\hat \sigma _y^{(1)}\otimes \hat \sigma _y^{(2)}$; then we have,
\begin{equation}
P_\mu \left( {B_j=\varepsilon _j\left|\,{\sigma _x^{(j)}=\varepsilon _j} 
\right.} \right)=1\,,\;\;\;\;\varepsilon _j=\pm 1\,,\;\;j=1,2,3\,,
\label{trece}
\end{equation}
\begin{equation}
P_\mu \left( {\sigma _x^{(1)}\sigma _x^{(2)}\sigma _x^{(3)}=-1} \right)=1\,.
\label{catorce}
\end{equation}

If the three $\hat \sigma _x^{(j)}$ are measured in spacelike separated 
regions of the same individual system in state (\ref{ocho}), equation 
(\ref{trece}) allows us to infer three joint SERs, $B_j=\varepsilon _j$, 
that must verify the relation $\varepsilon _1\varepsilon _2\varepsilon _3=-1$ 
following equation (\ref{catorce}). On the other hand, the product of the 
three {\em compatible} observables $\hat B _{j}$ is the unit operator, and 
therefore there is {\em no} quantum state in which the results of their 
measurement in the same individual system satisfy that relation 
(each result will be $\pm 1$, but the product of the three is always $+1$). 
This proves again the incompatibility between QM and joint SERs.

Note that all observables used to infer ERs in sections 4 and 5 are (local) 
spin components, and therefore there would be no difficulty for their joint 
measurement and the corresponding inference of joint SERs. Non-localities 
appear only in the observables to which the ERs are assigned; in this sense 
the experimental check of some of the predictions in sections 4 and 5 could 
be problematic, but our (gedanken) examples show the differences between QM 
{\em theory} and any theory that includes EPR's elements of reality: independently 
of any experimental confirmation of either theory, we have shown that ERs 
cannot be used to ``complete'' QM.

\section{Summary}
Working only with what we have called ``joint SERs'', inferred from joint 
measurements in the same individual physical system 
(a more {\em actual} kind than 
the usual ERs inferred from {\em alternative} incompatible measurements), 
we have 
reached the following conclusions.

(i) If we accept a very mild sufficient condition for the existence 
of elements of reality, joint SERs for observables without {\em any} common 
eigenstate will exist. On these premises, QM would be incomplete. 
This formulation of the EPR argument in terms of joint SERs was considered 
impossible.

(ii) There are sets of joint SERs for compatible observables that, 
according to QM, can never be obtained as results of the corresponding joint
measurements in {\em any} individual system. However plausible, elements of 
reality are incompatible with QM (Bell's theorem).

Both theorems have been proved using quite similar arguments in the same 
{\em entangled} quantum states (\ref{uno}), (\ref{ocho}).

\section{Acknowledgements}
We would like to thank Gabriel \'{A}lvarez and Emilio Santos for their many 
helpful comments and the anonymous referees for pointing out to us some 
obscure points and for their constructive comments.



\begin{thebibliography}{99}
\bibitem{EPR35} Einstein A, Podolsky B and Rosen N 1935 {\em Phys. Rev.} 
{\bf 47} 777
\bibitem{Hardy93} Hardy L 1993 {\em Phys. Rev. Lett.} {\bf 71} 1665
\bibitem{GHZ89} Greenberger D M, Horne M A and Zeilinger A 1989 
{\em Bell's Theorem, Quantum	Theory, and Conceptions of the Universe} ed
M Kafatos (Dordrecht: Kluwer) p 69
\bibitem{Mermin90a} Mermin N D 1990 {\em Phys. Rev. Lett.} {\bf 65} 3373
\bibitem{Mermin90b} Mermin N D 1990 {\em Boojums all the Way Through} 
(Cambridge: Cambridge	University Press) p 177
\bibitem{HB92}	H\'{a}jek A and Bub J 1992 {\em Found. Phys.} {\bf 22} 313
\bibitem{Howard85}	Howard D 1985 {\em Stud. Hist. Phil. Sci.} {\bf 16} 171
\bibitem{Redhead87} Redhead M L G 1987 {\em Incompleteness, Nonlocality, 
and Realism} (Oxford:	Clarendon)
\bibitem{Ballentine90} Ballentine L E 1990 {\em Quantum Mechanics} 
(Englewood Cliffs, NJ: Prentice-Hall)		p 453
\bibitem{CS78} Clauser J F and Shimony A 1978 {\em Rep. Prog. Phys.} 
{\bf 41} 1881
\bibitem{GHSZ90} Greenberger D M, Horne M A, Shimony A and Zeilinger A 1990 
{\em Am. J. Phys.}	{\bf 5} 1131
\bibitem{Stapp91} Stapp H P 1991 {\em Found. Phys.} {\bf 21} 1
\bibitem{d'Espagnat93} d'Espagnat B 1993 {\em Bell's Theorem and the 
Foundations of Modern Physics}	ed A van der Merwe and F Selleri 
(Singapore: World Scientific) p 139
\bibitem{Shimony93} Shimony A 1993 {\em Search for a Naturalistic World View} 
vol II (New York: Cambridge	University Press) p 187
\bibitem{Pitowsky91} Pitowsky I 1991 {\em Phys. Lett. A} {\bf 156} 137
\bibitem{BH93} Bohm D and Hiley B J 1993 {\em The Undivided Universe} 
(London: Routledge)
\bibitem{Bell64} Bell J S 1964 {\em Physics} {\bf 1} 195
\end{thebibliography}
\end{document}